\documentclass[12pt]{article}

\usepackage{graphicx}
\usepackage[]{hyperref}
\usepackage{epstopdf}

\usepackage{color,multicol}
\usepackage{amsmath}
\usepackage{amssymb}
\usepackage{pifont}
\usepackage{float}
\usepackage{cite}

\usepackage{tocloft}

\begin{document}


\title{{\large ONE TRADE AT A TIME --\\ UNRAVELING THE EQUITY PREMIUM PUZZLE}\footnote{Original version 25 July 2015. {\sl Journal-ref:} Supplementary materials for \lq\lq \href{https://www.mdpi.com/1099-4300/22/8/860}{Economics of Disagreement} -- financial intuition for the R\'enyi divergence'', Entropy {\bf 22}(8), 860 (2020).\vspace*{1mm}}}
\author{Andrei N. Soklakov\footnote{
Head of Strategic Development, Asia-Pacific Equities, Deutsche Bank.\vspace*{1mm}\newline
{\sl The views expressed herein should not be considered as investment advice or promotion. They represent personal research of the author and do not necessarily reflect the view of his employers, or their associates or affiliates.} Andrei.Soklakov@(db.com, gmail.com).}}
\date{}


\maketitle

\vspace*{-0.5cm}

\begin{center}
\parbox{14cm}{
{\small
Financial markets provide a natural quantitative lab for understanding some of the most advanced human behaviours. Among them is the use of mathematical tools known as financial instruments. Besides money, the two most fundamental financial instruments are bonds and equities. More than 30 years ago Mehra and Prescott found the numerical performance of equities relative to government bonds could not be explained by consumption-based (mainstream) economic theories. This empirical observation, known as the Equity Premium Puzzle, has been defying mainstream economics ever since. The recent financial crisis revealed an even deeper need for understanding financial products. We show how understanding the rational nature of product design resolves the Equity Premium Puzzle. In doing so we obtain an experimentally tested theory of product design.\\

\phantom{{\sl Keywords:} }risk aversion.\\


}
}
\end{center}

\newpage

\section{Introduction and background}

Following the latest crisis, the economics community as well as the entire financial industry attracted a new wave of criticism. This time, a large portion of the blame fell on financial products. The quality and even the purpose of financial products were questioned.

Critical review of financial instruments has led us to a theory of rational product design~\cite{Soklakov_2014_WQS}.
Within this theory the mathematical structure of a financial product is no longer left to intuition. It follows logically from the customer's intended behaviour.

In science, theories gain credibility by surviving tough existential tests against reality. The infamous Equity Premium Puzzle~\cite{MehraPrescott_1985} provides us with a perfect challenge. It comes with a rich independent history (defying mainstream economics for over 30 years) and, at the same time, it is intimately linked to financial products. Indeed, how can anyone claim to understand financial products without understanding the real-life performance of such basic instruments as equities and bonds?

We find our approach (let us call it \lq\lq Quantitative Structuring") correctly predicts the observed data. Remarkably, even if the numerical values for the equity premium were somehow not known, our approach would have predicted the right ballpark range.

Quantitative Structuring does not follow the classical paradigm of a consumption-based economy populated by identical copies of a representative agent. Instead, it advocates a detailed naturalistic view of economics with large behavioural diversity including some behaviours that are not shaped by consumption.

The evolutionary aspect of our approach has counterparts in both evolutionary economics and evolutionary finance (see e.g. \cite{NelsonWinter_2002,EvstigneevEtAl_2009} and references therein). Deep down, however, our approach is a lot less ambitious: it really is just a rational framework for manufacturing financial products for the benefit of individual customers. The evolutionary elements of our approach are not postulated, they appear naturally as a consequence of rational product design: products which make no rational sense do not survive.

Regarding the concept of consumption, we do not doubt its importance in economics. However, we do challenge its explanatory power at the level of individual strategies. In this regard we are similar to a naturalist who does not doubt the consumption of solar power as a major factor driving life on Earth. However, the naturalist would never support a theory which tries to reduce every observation to optimal consumption of solar power by a \lq\lq representative life-form''.
The economic landscape is populated by a diverse set of strategies~\cite{Damodaran_2012}. Each strategy is rational in its own way and leaves behind its own mark. The data on aggregate consumption, let alone the concept of a representative agent, are too high-level to appreciate every rational consequence within this diverse reality.

Speaking about diversity of financial strategies we must, at the very least, distinguish between investing and hedging. Traditional economic models tend to focus on hedging (even when discussing investments). Indeed, models which assume market efficiency can do little else. Quantitative Structuring complements this by describing a class of pure investment strategies where the investor earns a living by having a view on the market (learning new information and bringing that information to the market).

The observed equity premiums appear to come from investment strategies and not from hedging behaviours such as smoothing of consumption or diversification.

\subsection{The Equity Premium Puzzle}
In 1985 Mehra and Prescott investigated historical data on the long-term excess returns achieved by equities over government bonds~\cite{MehraPrescott_1985}. These excess returns, known as the equity premium, appeared to be surprisingly high (confirming an independent observation of Shiller~\cite{Shiller_1982}). Mehra and Prescott estimated that the equity premium was an order of magnitude greater than could be rationalized within the standard consumption-based theories.

Given the importance of this challenge, proposals to resolve the puzzle quickly snowballed. Two decades later Mehra and Prescott revisited the progress on the problem only to reinforce their original conclusions~\cite{MehraPrescott_2003}. They estimated the equity premium to be up to 8\% in arithmetic terms or up to 6\% in terms of geometric (compound) returns and reiterated the Equity Premium Puzzle as an unresolved challenge to explain these values. A more recent independent study suggested that the puzzle may be even deeper than originally estimated~\cite{Azeredo_2014}. At the time of writing it would be fair to say that no single explanation of the puzzle has yet received general acceptance and the search for a clear dominant explanation continues.

So what exactly is the problem? There are two main aspects to the puzzle. First, we need to find the underlying mechanism which explains the observed numbers (the recorded premiums of up to 6\% annualized compound returns). This is our primary goal. The second (more contentious) aspect is the very practice of science in economics. Our goal here is not to complicate a theory until it accommodates the possibility of high equity premiums. What we need is an independently motivated (i.e. not specialized around equity premiums) {\it parsimonious} theory which {\it predicts} the correct magnitude of the premium.

In their original paper~\cite{MehraPrescott_1985} Mehra and Prescott considered the classical model based on two fundamental assumptions. The first assumption states that the entire economy consists of very similar participants; so much so that it can be described using the concept of a representative agent. The second assumption states that the agents' motivation can be summarized as consumption optimization. This model is a perfect example of parsimony. It makes very specific testable predictions about the equity premium. Unfortunately, as shown by Mehra and Presott, the predictions disagree with the reality~\cite{MehraPrescott_1985}.

Mehra and Prescott put the classical assumptions into question. The representative agent model was challenged almost immediately after the discovery of the puzzle~\cite{Abel_1988}. Despite becoming somewhat of an ideology, the postulate of optimal consumption also started to lose ground as behavioural corrections were introduced \cite{Abel_1990, BenartziThaler_1995}.

Unfortunately, the debates were set to continue as new developments appeared at the expense of parsimony. Most new theories maintained significant allegiance to the original consumption-based theory, upgrading it with additional factors rather than proposing a genuine alternative. For example, the myopic loss aversion effect in~\cite{BenartziThaler_1995} can be introduced, removed or adjusted at will depending on the frequency of portfolio evaluations. Similarly, the habit formation models (see e.g. \cite{CampbellCochrane_1999}) include the classic model of optimal consumption as a special case. Such mathematical anchoring on the idea of optimal consumption meant that, in contrast to the actual observations, the new theories have never really rejected small equity premiums. They simply introduced additional parameters to accommodate a wider range of possibilities.

Could it be that not all economically significant human behaviours optimize consumption? In particular, could it be that consumption optimization does not cause the observed equity premiums? We think that the answer for both of these questions is yes. The long history of the Equity Premium Puzzle justifies considering such a possibility. In this paper we refrain from using both the idea of a representative agent and the postulate of optimal consumption.

Of course, we cannot hope to build a more accurate theory by simply removing some old assumptions. We need an alternative theory.
The principle of parsimony is a very tough epistemological principle which sets very high standards. At the time of writing, we are not aware of any independently motivated fully heterogeneous non-consumption-based models in the literature on the equity premium puzzle that accurately predict the premiums without opening themselves to the criticism of insufficient parsimony. This brings us to the topic of the next section.

\subsection{Financial products}
Financial products survive by making sense. Making sense of the equity premium is best approached as part of a larger goal -- understanding equities as a financial product. In this section we introduce a theory of rational product design which helps to achieve that.

\subsubsection{Quantitative structuring (general motivation)}

Financial products are defined by their payoff function, $F(x)$, which states how the benefits (usually cash flows) depend on the underlying variables, $x$. Such products, also known as {\it financial derivatives}, are designed to be bought and sold as a single trade.

We have been trading financial products since ancient times. However, it is only relatively recently, building on the works of Black, Scholes and Merton~\cite{BlackScholes_1973, Merton_1973}, that we found a rational framework for pricing such products. In order to price a product, defined by its payoff $F$, we compute a quantity of the form
\begin{equation}\label{Eq:GeneralPricing}
{\rm Price}(F\,)=\sum_x F(x)\,Q(x)\,,
\end{equation}
where the summation is taken over all possible values of the underlying variables and where the weighting function $Q$ is implied by a so-called pricing model.

Historically, within the industry of financial derivatives, most quantitative effort went into modeling, i.e.\/ the study of $Q$. This was enough to create a dedicated mathematical discipline and an entire new profession of \lq\lq financial engineers'' better known as \lq\lq quants''. Quants were not normally involved in product design. \lq\lq Good quants" were expected to price any product. For them products were challenges, a source of modeling requirements.

A closer look at Eq.~(\ref{Eq:GeneralPricing}) reveals that the value of a product is determined by both its payoff structure $F$ and the pricing model $Q$ in a remarkably {\it symmetric} way. If modeling requires a dedicated mathematical framework, so does product design.

Quantitative Structuring is a technical framework for financial product design~\cite{Soklakov_2014_WQS}. It can be described as a fusion between basic logical elements of learning theory and rational optimization. The following subsections provide a brief technical introduction.

\subsubsection{Information derivatives}

Consider an investor with an interest in some economically-relevant variable $x$ (future stock price, currency exchange rate, temperature readings, etc.). The investor's entire knowledge about the variable can be summarized as a list of all possible values for the variable alongside the investor's opinion on their probabilities $\{x,p(x)\}$. We recognize this as a probability distribution and conclude that the investor's knowledge should be subject to the logic of probability theory. In particular, upon discovery of some data $d$ any logical investor must update their knowledge from $p(x)$ to $p(x|d)$ using Bayes' theorem
\begin{equation}\label{Eq:Bayes}
p(x|d)={\cal L}_d(x)\,p(x)\,,
\end{equation}
where ${\cal L}_d(x)$ is known as the likelihood function.

Let the probability distribution $p(x)$ summarize all known information about $x$. We see that the likelihood function ${\cal L}_d(x)$ in Eq.~(\ref{Eq:Bayes}) is the only object describing the learning of new information on top of the already known $p(x)$. If we believe in research-based investments which deliver new information to the markets, we must seek a framework in which the payoff function $F(x)$ of any $x$-contingent investment is explicitly related to the likelihood function ${\cal L}_d(x)$ describing the relevant research.

We use the term {\it information derivatives} to describe financial products whose structure is explicitly derived from all the relevant information. Ultimately, all financial products must evolve to satisfy this definition. In the next subsection we go deeper into the mathematical structure of information derivatives.

\subsubsection{Structuring equations}

It would be useful to find the framework of information derivatives within the established (and much more general) paradigm of rational optimizations. That would also clarify the goals achieved by financial products.

To our knowledge, Bernoulli's treatment of the St Petersburg paradox~\cite{Bernoulli_1738} was the first example of understanding investors' rationale in the context of a single financial product. The product was entirely hypothetical (the payoff from a certain game) but, never\-theless, it had the defining ingredient of all modern financial derivatives -- a contingent cashflow deriving its value from an outcome of a random variable.

Von~Neumann and Morgenstern put Bernoulli's intuitive understanding of rationality on firm axiomatic grounds~\cite{NeumannMorgenstern_1944}. The expected utility framework which followed from this work turned out to be very rich, containing many models including ones that have no meaningful connection to the process of learning as described by Eq.~(\ref{Eq:Bayes}). For instance, it contains the famous Capital Asset Pricing Model (CAPM). Basic CAPM (see e.g.~\cite{Sharpe_1964}) assumes that all information is completely objective and is agreed upon by all investors, so there is literally nothing left to learn.

Let us come back to the original Bernoulli setup and allow the investor complete freedom to develop their knowledge (both about the market and about their private circumstances). The optimal payoff function $F$ for such an investor would solve an optimization problem of the standard form
\begin{equation}\label{Eq:Optimization}
\max_F\int b(x)\, U(F(x))\,dx\ \ \ \ {\rm subject\ to\ budget\ constraint}\ \ \ \ \int F(x)\, m(x)\,dx=1\,,
\end{equation}
where $U$ is the utility function, $b(x)$ is the {\it investor-believed} probability distribution, and the values of $\{m(x)\}$ are offer prices for generalized Arrow-Debrew securities (these can be readily obtained for almost any variable by calling an investment bank). After a trivial normalization, $\{m(x)\}$ defines a probability distribution which we call {\it market-implied}.

It turns out that the solution $F$ of the optimization~(\ref{Eq:Optimization}) can be understood in terms of the likelihood functions describing investment research -- just as we demanded above. In particular, we can imagine two learning steps of the Bayesian form~(\ref{Eq:Bayes}). One of these steps refers to the discovery of the investor-believed probabilities about the underlying variable, and the other describes a much more private learning of the investor's own preferences (risk aversion). The two steps correspond to a pair of equations (see~Ref.~\cite{Soklakov_2013b} for details)
\begin{equation}\label{Eq:b=fm}
b=f\,m
\end{equation}
\vspace*{-5mm}
\begin{equation}\label{Eq:PayoffElasticity}
\frac{d\,\ln F}{d\,\ln f}=\frac{1}{R}\,,
\end{equation}
where the risk aversion coefficient $R$ is connected to the utility $U(F)$ through the standard Arrow-Pratt formula: $R=-FU''_{FF}/U'_F$, and $f(x)$ can be understood as the payoff function for the growth-optimal product ($F=f$ when $R=1$ -- the case of a Kelly investor~\cite{Kelly_1956}).

The logical structure~(\ref{Eq:Bayes}) of Eq.~(\ref{Eq:b=fm}) is easy to see: the market-implied $m$ and the investor-believed $b$ naturally take the places of the prior and the posterior distributions respectively, while the growth-optimal payoff $f$ coincides with the likelihood function.

Understanding more general payoffs in terms of the relevant likelihood functions and how that leads to Eq.~(\ref{Eq:PayoffElasticity}) is a bit more involved. For more detailed explanations of Eqs.~(\ref{Eq:b=fm}) and~(\ref{Eq:PayoffElasticity}) including motivation, derivations, intuitive illustrations as well as concrete numerical examples (including product performance), we refer the reader to~\cite{Soklakov_2014_WQS,Soklakov_2013b,Soklakov_2011,Soklakov_2013a,Soklakov_2014MR} and \cite{Soklakov_2018}.

\subsubsection{Comparison to consumption optimization}

The expected utility framework is so good at unifying diverse economic motivations that, looking at Eq.(\ref{Eq:Optimization}), it may not be obvious how Quantitative Structuring is different from the consumption-based approach that is used in the literature on the equity premium puzzle. To see this more clearly, consider an investor which agrees with the market. Substituting $b(x)=m(x)$ into Eq.~(\ref{Eq:b=fm}) we see that the investor would not trade $x$ (the payoff does not depend on $x$). In real life such an investor would say that they \lq\lq have no edge" or \lq\lq see no investable opportunity" on the market. Equations~(\ref{Eq:b=fm}) and~(\ref{Eq:PayoffElasticity}) convert learning into product design. No original view on the market means no trade.

Contrast this with the setup in Mehra and Prescott~\cite{MehraPrescott_1985} where none of the agents have any view on the market. For them the value of a financial product comes from its covariance with consumption. The strategy of such agents is best described as hedging (smoothing of consumption) rather than investing.

\section{Results}
\subsection{Expected premiums}
The Equity Premium Puzzle is about the already {\it realized} performance of equities above bonds. However, the realized performance is only half the picture. Investors do have expectations. These expectations must make sense. Without that the investments have little chance of even starting, let alone surviving long term. We must understand the investor-expected premiums.

\begin{figure}[t!]
\vspace*{-0.5cm}
\hspace*{-0.12\textwidth}\includegraphics[width=1.24\textwidth]{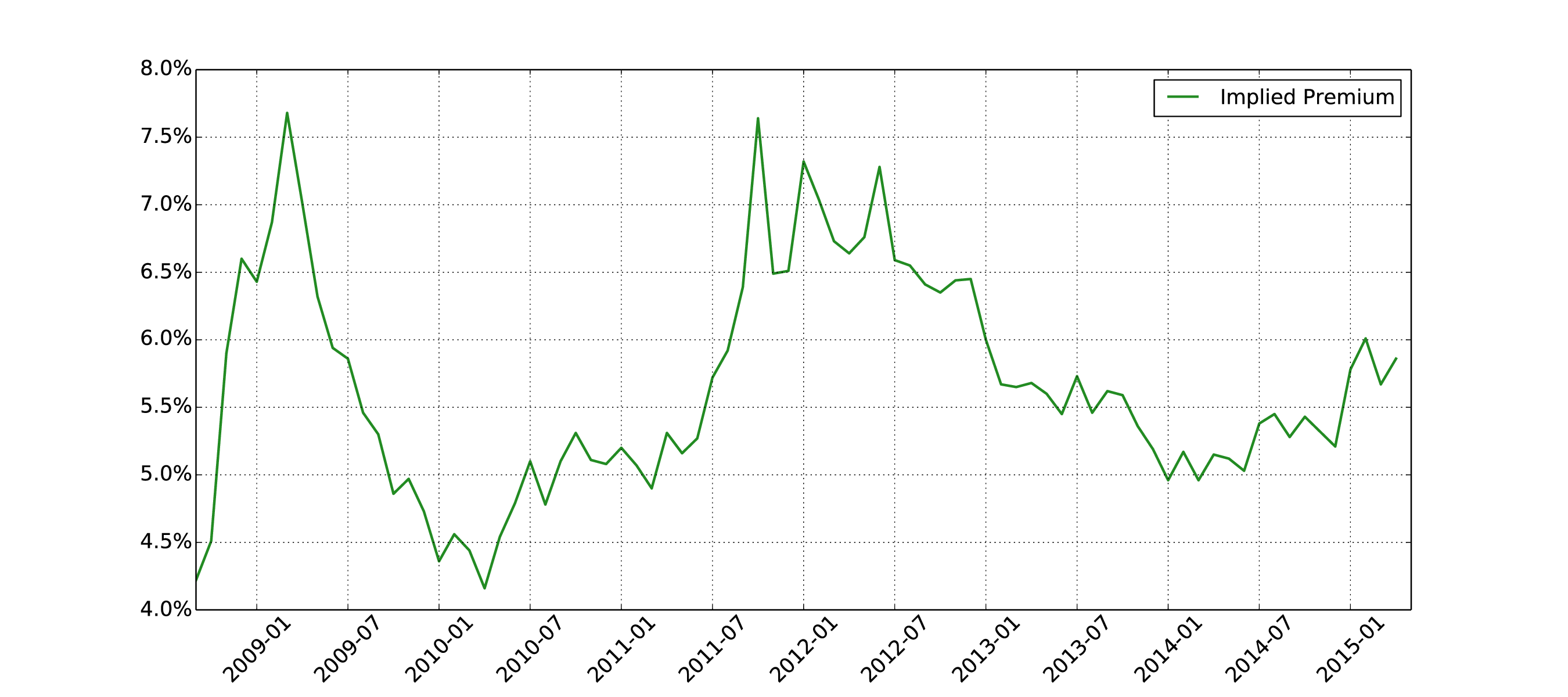}
\caption{Implied Equity Premiums (annualized compounded) from Damodaran~\cite{Damodaran_2014}. The records are updated on a monthly basis starting from September 2008. The last data point on the graph is dated April 2015. The values are quoted at the beginning of each month.  In our calculations we interpreted this as the first business day of each month.\newline }\label{Fig:ImpliedPremium}
\end{figure}

On Fig.~\ref{Fig:ImpliedPremium} we display independent empirical quotes for the expected equity premiums as reported by Damodaran~\cite{Damodaran_2014} using SPX data. We see these values are just as large as the historical records by Mehra and Prescott -- at least an order of magnitude above 0.35\% per annum quoted in~\cite{MehraPrescott_1985}. The aim of this section is to explain these values.

Using the notation of~(\ref{Eq:Optimization}), we can write the investor-expected continuously-compounded rate of return as
\begin{equation}\label{Eq:ER}
{\rm ER}=\int b(x) \ln F(x) \,dx\,.
\end{equation}
Let us choose $x$ to be the total return on one unit of wealth invested in some equity. With this choice of $x$ the payoff of the equity investment is simply
\begin{equation}
F(x)=x\,.
\end{equation}

Mehra and Prescott considered an investor with constant relative risk aversion, $R$. In this case Eq.~(\ref{Eq:ER}) becomes (see Eq.~(\ref{Eq:ER^Z_R}) in Appendix~\ref{Sec:ER_R ER_R^LN})
\begin{equation}\label{Eq:ER_R}
{\rm ER}_{R}=\frac{1}{{\rm Price}(x^R)}\frac{\partial {\rm Price}\big(x^R\big)}{\partial R}\,.
\end{equation}
In order to compute this quantity we just need the ability to price power payoffs, $x^R$. For that we use the industry-standard approach of static replication with vanilla options (see Ref.~\cite{CarrMadan_2001} and Appendix~\ref{Sec:ER_R}). In terms of market information, this replication needs the implied volatility curve at the same maturity as the payoff $x^R$.

\begin{figure}[t!]
\vspace*{-0.5cm}
\hspace*{-0.12\textwidth}\includegraphics[width=1.24\textwidth]{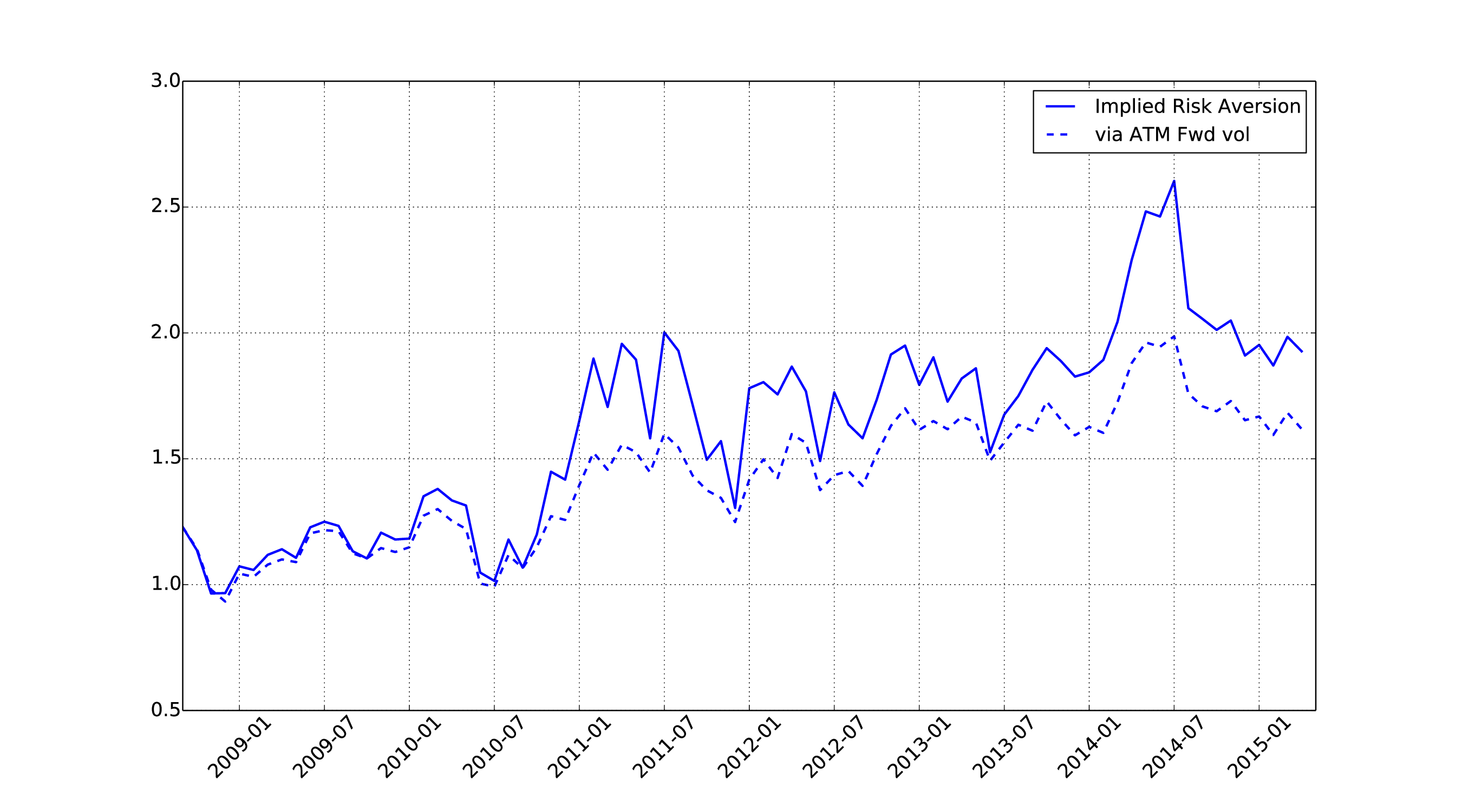}
\caption{Implied risk aversion is comfortably below 3. Solid and dashed lines show the value of $R$ obtained by reconciling Eq.~(\ref{Eq:ER_R}) and Eq.~(\ref{Eq:ER_R^LN}) respectively to the data of Fig.~\ref{Fig:ImpliedPremium}.
}\label{Fig:Rimp}
\end{figure}

According to Damodaran~\cite{Damodaran_2014}, his quotes for the premiums accurately reflect detailed market information (such as market-implied dividends) of up to five years into the future. Similarly, our SPX volatility surfaces use exchange traded options with maturities of up to five years. The solid line on Fig.~\ref{Fig:Rimp} depicts the values of $R$ obtained from reconciling Eq.~(\ref{Eq:ER_R}) with the data on Fig.~\ref{Fig:ImpliedPremium} using the complete historical records of 5-year volatility curves.

In order to verify the robustness of our calculations, reduce the dependency on market data sources, and to develop a more transparent example of our calculations we also consider a flat-volatility market
\begin{equation}\label{Eq:LN_distr}
m(x)=\frac{\rm DF}{x\sigma\sqrt{2\pi}}\exp\Big{\{}-\frac{(\ln x-\mu)^2}{2\sigma^2}\Big{\}}\,,\ \ \mu=r-\sigma^2/2\,,
\end{equation}
where ${\rm DF}$ is the discount factor, $r$ is the risk free return and $\sigma$ is the volatility.
In this case ${\rm ER}_{R}$ is reduced to a simple analytic expression (see Eq.~(\ref{Eq:appendixLN}) in Appendix~\ref{Sec:ER_R ER_R^LN}):
\begin{equation}\label{Eq:ER_R^LN}
{\rm ER}_{R}\to{\rm ER}_{\rm R}^{\rm LN}=r+(R-1/2)\sigma^2\,.
\end{equation}
This gives us the ballpark estimate of the expected equity premium of $(R-1/2)\sigma^2$.

\begin{figure}[t!]
\vspace*{-0.5cm}
\hspace*{-0.12\textwidth}\includegraphics[width=1.24\textwidth]{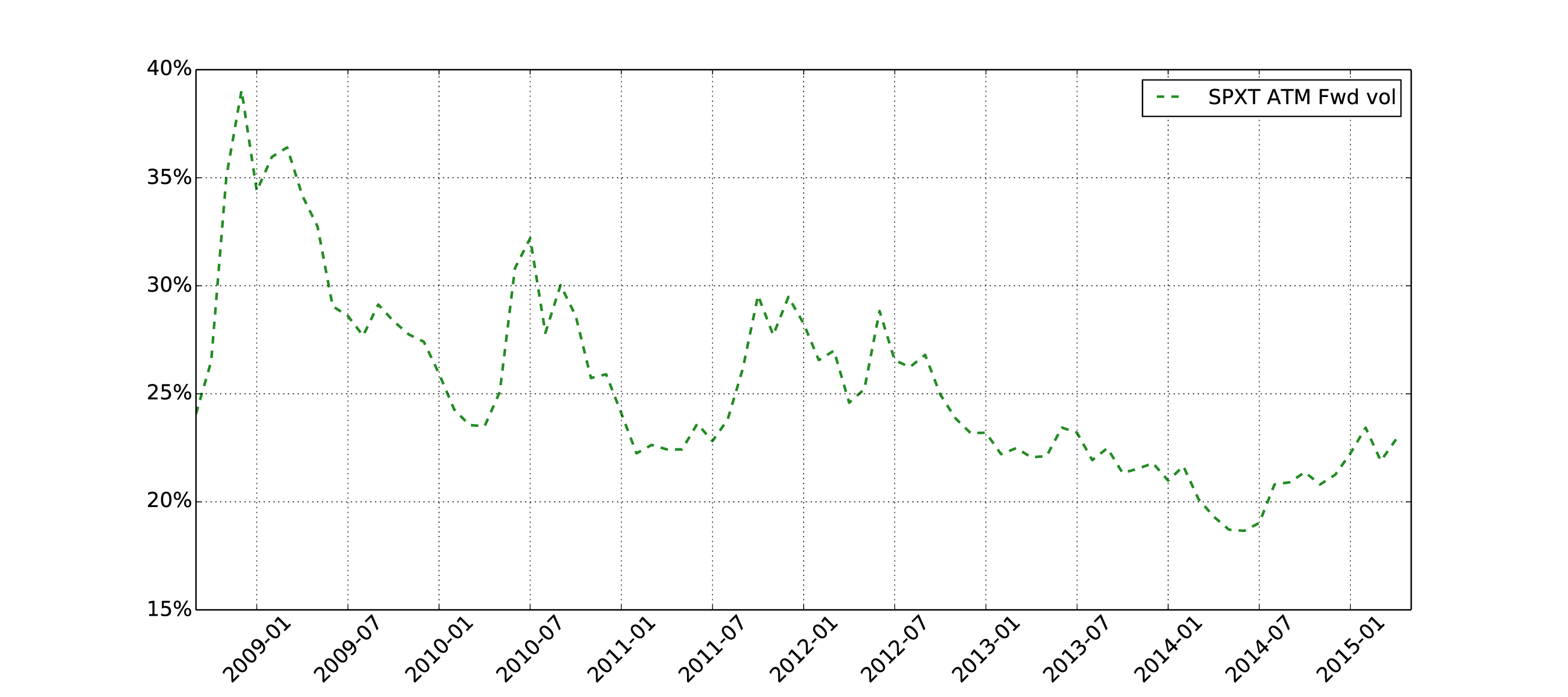}
\caption{SPXT 5-year at-the-money-forward values of implied volatility (annualized).\newline }\label{Fig:SPXT_ATMF_Vol}
\end{figure}

The dashed line on Fig.~\ref{Fig:Rimp} shows the value of $R$ implied from equating the annualized value of the premium $({\rm ER}_{\rm R}^{\rm LN}-r)$ with the relevant quoted value from Fig.~\ref{Fig:ImpliedPremium}. In this calculation we used the 5-year at-the-money-forward implied volatilities (displayed for convenience on Fig.~\ref{Fig:SPXT_ATMF_Vol}). Both graphs on Fig.~\ref{Fig:Rimp} show good agreement indicating robustness of our calculations.

In their pioneering paper~\cite{MehraPrescott_1985}, Mehra and Prescott argue that the acceptable values for $R$ must be below 10. In fact, all of the actual estimates of $R$ which they cite to support their argument were below 3.\footnote{Risk aversion data often come from experiments in which people are presented with a single trade much like in Bernoulli's original research on the St Petersburg paradox. One can argue that all subsequent interpretation of such data should be made using a similar {\it single-trade} perspective.} This is in remarkable agreement with Fig.~\ref{Fig:Rimp}.

Even with minimal knowledge of volatility, making the standard assumption of 20\% for typical equity volatility, the risk aversion values of up to 3 allow us to explain premiums as high as 10\% in terms of continuously compounded annual returns.

We conclude that, in terms of investors' expectations, Quantitative Structuring is consistent with the observed equity premiums.\\

\subsection{Realized premiums}

In the above section we presented a rational explanation for the investor-expected equity premiums. We also noticed that the investor-expected premiums fall in the same ballpark range as the actually realized premiums. In this section we would like to understand how this happens: how the investors' expectations materialize, with investors doing no more than just keeping their money in the equity.

Let $S_t$ be the value of the total return version of some equity index at time $t$. The return on the equity can be partitioned arbitrarily into $N$ imaginary reinvestment steps:
\begin{equation}\label{Eq:ReinvestmentSteps}
S_N=S_0\cdot\frac{S_1}{S_0}\cdot\frac{S_2}{S_1}\cdots\frac{S_N}{S_{N-1}}\,.
\end{equation}
Defining $x_i=S_i/S_{i-1}$ we compute
\begin{equation}
S_N=S_0\prod_{i=1}^N x_i= S_0e^{\sum_{i=1}^N\ln x_i}=S_0e^{N \cdot {\rm Rate}}\,,
\end{equation}
where
\begin{equation}\label{Eq:HistRate}
{\rm Rate}=\frac{1}{N}\sum_{i=1}^N\ln x_i\,.
\end{equation}
Let us now look at this quantity using the standard statistical approach. In this approach the individual elements $\{x_i\}$ are viewed as realizations of a random variable $X$ with some (possibly unknown) distribution $P(X)$. For the basic statistical concepts to make practical sense, the law of large numbers is assumed to hold.\footnote{This is true if the individual elements, $\{x_i\}$, are sufficiently independent from each other.} In this framework, as $N$ increases, the average~(\ref{Eq:HistRate}) converges almost surely to the expectation:
\begin{equation}\label{Eq:LawLargeNumbers}
{\rm Rate}\stackrel{\rm a.s.}{\longrightarrow}\int P(x) \ln x\, dx\,.
\end{equation}
Compare this with Eq.~(\ref{Eq:ER}) (remember $F(x)=x$ for equity investments). We see that the investor-expected returns are achieved when the investors' market beliefs agree with the reality, i.e. when $b=P$. Correct beliefs is indeed a very natural condition for the realized returns to agree with the expectations.

It turns out that the same condition, $b=P$, determines the level of risk aversion that the equity investor must have. Indeed, substitution of both $F$ and $m$ into Eqs.~(\ref{Eq:b=fm}-\ref{Eq:PayoffElasticity}) leaves us with the connection between $b$ and $R$. Setting $b=P$ determines the value of $R$. We shall
refer to such values of $R$ using the words \lq\lq historical" or \lq\lq realized" (as they are implied from the distribution $P$ of the realized returns). Following Mehra and Prescott~\cite{MehraPrescott_1985}, we want to compute these historical values of risk aversion and check if they are realistic.

Before we do that, let us briefly discuss the choice of partitioning in Eq.~(\ref{Eq:ReinvestmentSteps}). Although we are free to examine any such partitioning, some choices are more interesting than others. Partitions with very small $N$ are not useful because they would not give us enough data to achieve statistically meaningful convergence~(\ref{Eq:LawLargeNumbers}). The opposite extreme of very large $N$ brings us to the domain of high-frequency trading which is normally practiced by people with a mindset that is very different from a long-term investment.

Ideally, we want to focus on the smallest possible $N$ that is large enough to ensure noticeable convergence~(\ref{Eq:LawLargeNumbers}). The standard deviation of the sum~(\ref{Eq:HistRate}) from its mean~(\ref{Eq:LawLargeNumbers}) scales as $N^{-1/2}$. For the first significant digit of the sum~(\ref{Eq:HistRate}) to emerge with reasonable probability, the convergence must reduce the standard deviation by an order of magnitude ($N^{-1/2}\sim 0.1$). This means that we must choose $N$ which is not much lower than 100.

We managed to find full market data, including volatility surfaces, for SPXT (total return version of SPX) going back to 17 May 2000. At the time of writing, this was about 15 years worth of data (daily records). Some researchers might argue the need for longer historical records. However, 15-year investments are already at the limit of what many people would consider practical, so we choose to accept it. Viewing 15 years of the entire investment history~(\ref{Eq:ReinvestmentSteps}) as if it was a sequence of bi-monthly reinvestments we get $N=90$ reinvestment periods. The corresponding returns $\{x_i\}_{i=1}^N$ can be used as Monte-Carlo realizations of $P(x)$. As discussed above, this amount of data is just enough to talk about averages like~(\ref{Eq:HistRate}) in terms of expectations~(\ref{Eq:LawLargeNumbers}).\footnote{Note, however, that the Monte-Carlo definition $\{x_i\}_{i=1}^N$ of $P$ ensures that the numerical computation of the expectations~(\ref{Eq:LawLargeNumbers}) coincides exactly with the realized rate of returns~(\ref{Eq:HistRate}).}

Using Eqs.~(\ref{Eq:b=fm}) and (\ref{Eq:PayoffElasticity}) and recalling for the simple equity investment $F(x)=x$, we compute
\begin{equation}\label{Eq:R}
    R=\frac{d\ln f}{d\ln F}=\frac{d\ln(b/m)}{d\ln x}=\frac{m}{b}\,\Big(\frac{b}{m}\Big)'_x\, x\,.
\end{equation}
Theoretically, this gives us the entire risk-aversion profile for the investor in question.

Right now, however, we have a bare minimum of statistical data. So, as many other researchers before us have done, we choose to focus on the overall level of risk aversion and defer the very interesting topic of the shape of risk-aversion profiles to further research. As a measure of the overall risk aversion we consider the average
\begin{equation}\label{Eq:ExpectedR}
\langle R\, \rangle_b \stackrel{\rm def}{=}\int R(x)\, b(x)\,dx\,.
\end{equation}
Put together, the above two equations give (see Eq.~(\ref{Eq:<R>b Analytic Appendix}) in Appendix~\ref{Sec:<R>b derivations})
\begin{equation}\label{Eq:<R>b Analytic}
\langle R\, \rangle_b =-1-\langle\, x (\ln m)'_x\,\rangle_b\,.
\end{equation}
This formula does not look very intuitive so, before using it, let us spend a few lines understanding it. To this end, let us see what it implies for a log-normal market-implied distribution. Substituting Eq.~(\ref{Eq:LN_distr}) into the above formula we derive (see Eq.~(\ref{Eq:<R>b Analytic explained Appendix}) in Appendix~\ref{Sec:<R>b derivations})
\begin{equation}\label{Eq:<R>b Analytic explained}
 \langle\,\ln x\,\rangle_b \ \stackrel{\rm LN}{=}  r+\Big(\langle R\, \rangle_b -1/2\Big)\sigma^2 \,.
\end{equation}
Compare this to Eq.~(\ref{Eq:ER_R^LN}) which we studied above. We recognize Eq.~(\ref{Eq:<R>b Analytic}) as a generalized analog of Eq.~(\ref{Eq:ER_R^LN}). The extent of generalization is substantial: the market can have any implied distribution, and the investor can have an arbitrary profile of risk-aversion.

We now recall taking the logical route where $b(x)=P(x)$ and, using Eq.~(\ref{Eq:<R>b Analytic}), we obtain the final formula for the overall level of risk aversion
\begin{equation}\label{Eq:<R>p}
\langle R\, \rangle_P =-1-\frac{1}{N}\sum_{i=1}^N x_i \Big(\ln m(x_i)\Big)'_{x_i}\,.
\end{equation}

\begin{figure}[t!]
\vspace*{-5mm}
\hspace*{-0.12\textwidth}\includegraphics[width=1.24\textwidth]{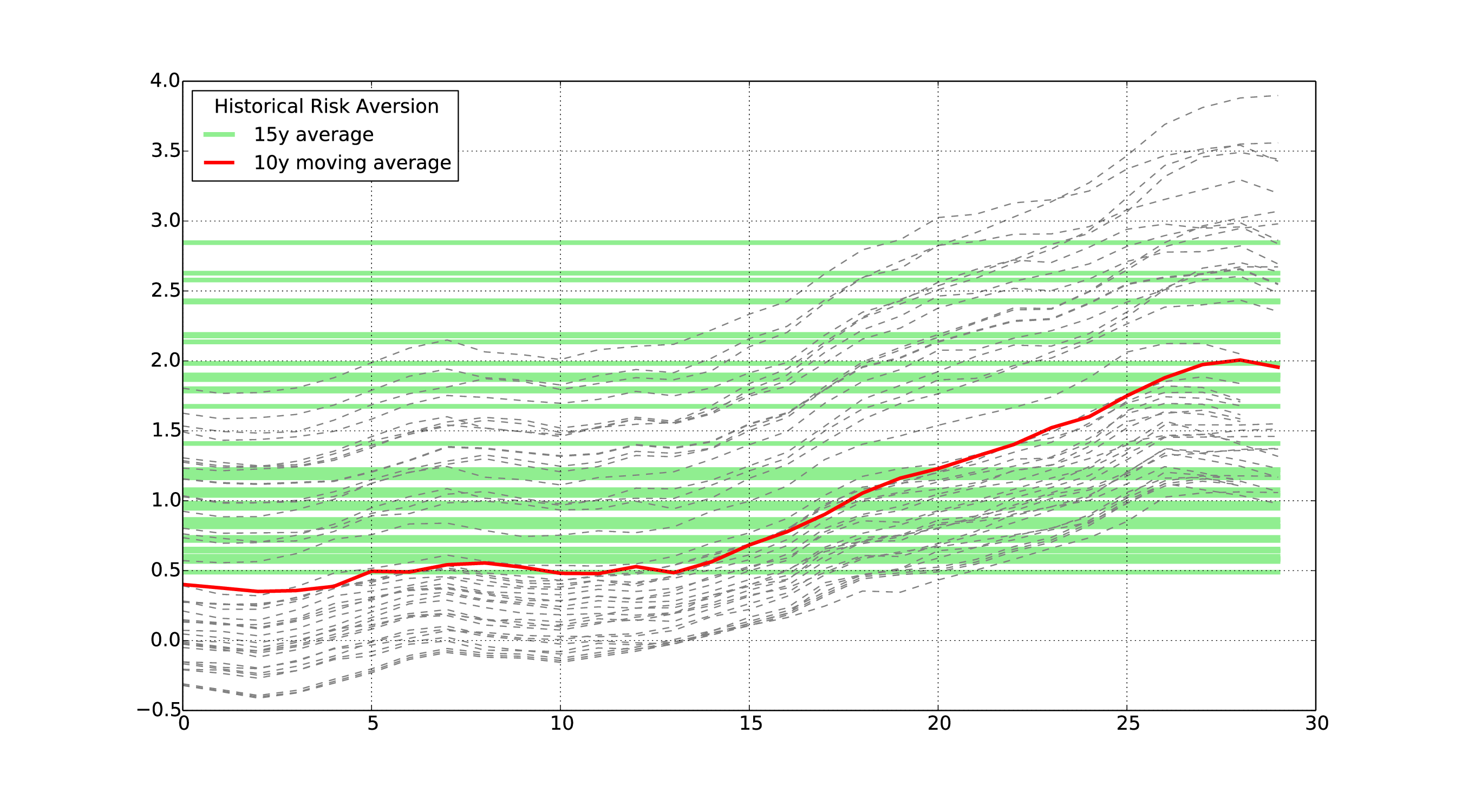}
\vspace*{-10mm}
\caption{Historical risk aversion. 10-year moving averages are computed on the bi-monthly grid as described in the main text. Within the 15-years of history this produces sequences of 30 (or 29) values (depending on the availability of data for the last period).\newline }\label{Fig:Rhist}
\end{figure}

We are now in a position to compute $\langle R\, \rangle_P$ as seen on any day for which we have enough market data to compute $m$ (see Appendix~\ref{Sec:<R>p}).

We should remember, however, that our investor took a 15-year view and is completely ignoring all intermediate updates from the markets. The level of risk aversion for such an investor should be measured in a way that represents most of the actual investment period and is not sensitive to intermediate market fluctuations. Below we report two sets of experiments which achieve this.

Before we describe the experiments, let us recall our choice to view historical investments as sequences of bi-monthly reinvestments. This has a useful side effect. A single experiment skips most of the available market data using only what it needs at bi-monthly intervals. The skipped market data can be used to repeat the experiment (42 times in total) -- we just need to start the bi-monthly sequence on a different business day within the first two months for which we have data.

In the first set of experiments we look at the averaged value of $\langle R\, \rangle_P$ across the entire 15-year investment period. These averaged values are reported on Fig.~\ref{Fig:Rhist} as the the horizontal (green) lines. Different lines correspond to the 42 different runs of the experiment. We see that all 42 levels of risk aversion are completely realistic. This is the main numerical result of this section and indeed the paper. We see that investors with totally ordinary levels of risk aversion achieve the observed equity premiums.

The second set of experiments has purely diagnostic purposes. We define these experiments by replacing the 15-year average with a 10-year moving average. The aim is to check the stability of our conclusions with respect to a significant variation of the setup. The moving average experiments also give us a glimpse of the term structure of risk aversion although, strictly speaking, such questions lie outside the scope of this paper.

The solid upward-trending (red) line on Fig.~\ref{Fig:Rhist} is a bi-monthly report of the 10-year moving average of $\langle R\, \rangle_P$ for the investment which started on the 17th of May 2000 -- the first day for which we have market data. The 42 runs of this experiment are plotted by faint dashed lines across the same graph. Although this experiment uses only 2/3 of the data (10 out of 15 years) and we can expect significant additional noise, we see that the risk aversion levels remain realistic. The increasing trend resonates with the expectations of Fig.~\ref{Fig:Rimp} and matches the increasingly risk averse sentiments on the global markets (from borderline careless prior the crisis to strongly risk averse after the crisis). These qualitative points support our confidence in the main results.

This concludes our analysis of the historical realized premiums. For additional technical notes and observations see Appendix~\ref{Sec:Misc}.

\section{Discussion}

Economics is an ecosystem populated by strategies. A single person can play host to many strategies and a single strategy can become popular among many people. Each strategy is rational. People, however, can make mistakes and may have complex personalities (inhabited by conflicting strategies). Advanced strategies rely, for their survival, on the mechanisms of learning. This puts pressure on the strategies to make sense, i.e. to offer understandable expectations which are confirmed by real-life performance.

In this paper we understood equity investments using a rational learning-based framework of Quantitative Structuring. We demonstrated in detail how investors with ordinary levels of risk aversion come to expect the seemingly high equity premiums and how such expectations materialize (and thereby confirm themselves) over long time horizons. In other words, our approach both predicts the correct value of the equity premium and provides a mechanism for its persistence in the real world.

The inability of consumption-based models to explain the observed equity premium (the Equity Premium Puzzle) implies that such models do not capture the relevant behaviour. In fact we see that the concept of consumption is not needed to explain the premium.

Consumption smoothing with little or no view on the market is best described as hedging. The observed equity premium is not caused by hedging. It is a signature of investments.

Quantitative explanation of the equity premium concludes the main part of this paper.

Along with equity investments, we could have just as easily examined any other investment strategy. Indeed, we could have studied virtually any payoff function $F$. It appears that we might have a general tool for examining the economic ecosystem at the level of individual strategies. Let us now conclude the paper by trying to understand the operation and the current limitations of such a tool.

As we have already mentioned, there is nothing new in trying to understand economics at a granular level. Most economists would already agree that the ideal theory should reflect the real diversity of strategies. The problem is that every attempt to get closer to this ideal inevitably faces the challenge of practicality. More detailed models need more detailed information.

Traditionally, the parsimony is enforced by making {\it ad hoc} simplifications: inventing representative agents, replacing the detailed description of an investment by a point on a mean-variance diagram. The resulting loss of information is hard to quantify and even harder to compensate for, even with the most reasonable of assumptions.

Information loss is bad for understanding investments because investments are extremely sensitive to information. Fortunately, investments themselves provide a lot of information. We see it reflected in their very structure -- the payoff function. The payoff function can appear very simple or very complex -- such appearances do not matter. What matters is that we know the {\it exact} payoff in {\it all} possible circumstances and that gives us a lot of information to work with.

The deep information content of payoff functions becomes both obvious and useful when we use it to derive the investors' views (see Eq.~(\ref{Eq:b=fm}) for a growth-optimizing investor or Eq.~(\ref{Eq:believed}) in Appendix~\ref{Sec:ER_R ER_R^LN} for a more general case). These views are very detailed -- they are complete probability distributions.

This is how we maintain the parsimony. On the one hand we treat investors as individuals who are free to learn and who are allowed to express any views they find logical. On the other hand we allow no room for speculation about what these views actually are. It is crucial that the views are not assumed, they are derived using the knowledge of payoff functions.

Discussions regarding the use of our methods in economics inevitably lead to the question of market equilibrium. Indeed, how could a model explain the existence of well-defined market prices if all investors within that model are allowed to have whatever views they want? It turns out the unique equilibrium market can be very easily derived (see Appendix~\ref{Sec:Equilibrium}).

It is important to remember that Quantitative Structuring has its own motivation which is quite separate from economics. The intrinsic motivation of Quantitative Structuring is the design of quality financial products. Independent motivation adds credibility to our analysis of the equity premium but it also comes with limitations. Eventually, Quantitative Structuring should cover both hedging and investments. However, the presented approach is currently limited to investments. Deep understanding of hedging strategies (deep enough to package them as financial products) is an important milestone for future work.

\begin{center}
    \rule[1ex]{.5\textwidth}{.5pt}\\[1cm]
\end{center}

\newpage

\setcounter{equation}{0}
\renewcommand\theequation{A\arabic{equation}}
\renewcommand{\thesubsection}{\Roman{subsection}}

{\Large\bf Appendix}
\subsection{Analytical expressions for ${\rm ER}_{R}$ and ${\rm ER}_{R}^{\rm LN}$} \label{Sec:ER_R ER_R^LN}
Equation~(\ref{Eq:PayoffElasticity}) can be rewritten as
\begin{equation}
d\ln f= R\,d\ln F\,.
\end{equation}
For the case of constant but otherwise arbitrary $R$ the above equation is immediately integrated to obtain
\begin{equation}
f(x)\propto e^{R\, \ln F(x)}=F^R(x)\,.
\end{equation}
This result together with Eq.~(\ref{Eq:b=fm}) give us the investor-believed distribution
\begin{eqnarray}\label{Eq:believed}
    b(x)&=&f(x)\,m(x)\cr
        &&\cr
        &=&\frac{e^{R\, \ln F(x)}m(x)}{\int e^{R\, \ln F(y)}m(y)\,dy}\,,
\end{eqnarray}
where we used the fact that $b(x)$ is normalized.
For the expected logarithmic return we compute
\begin{equation}
    {\rm ER}_{R} = \int b(x)\, \ln F(x) \,dx
                 = \frac{1}{Z}\,\frac{\partial Z}{\partial R}\,,\label{Eq:ER^Z_R}
\end{equation}
where
\begin{equation}
Z=\int F^R(x)\,m(x)\,dx\,.
\end{equation}
In this paper we focus on a straightforward equity investment. In this case $F(x)=x$, and $Z$ becomes essentially the $R$th moment of $m$. In the special case of log-normal market-implied distribution, this can be computed analytically (see Eq.~(\ref{Eq:LN_distr}) for notation)
\begin{equation}
Z=\int x^R\,m(x)\,dx={\rm DF}\cdot\exp\Big{\{}R\mu+\frac{1}{2}\,R^2\sigma^2\Big{\}}\,,
\end{equation}
and, in this special case, ${\rm ER}_{R}$ becomes
\begin{equation}\label{Eq:appendixLN}
{\rm ER}_{R}\to{\rm ER}_{R}^{\rm LN}=\mu+R\sigma^2\,.
\end{equation}

\subsection{Numerical computation of ${\rm ER}_R$}\label{Sec:ER_R}
According to Eq.~(\ref{Eq:ER_R}), we need the ability to compute ${\rm Price}(x^R)$. It is helpful to start with a more general case: instead of $x^R$, let us consider a twice-differentiable function $g(x)$. Using the popular notation $(\cdot)^{+}=\max(\cdot,0)$ we can write $g(x)$ in a form that is well known in financial applications (see Eq.~(1) in Ref.~\cite{CarrMadan_2001})
\begin{equation}
g(x)=g(k)+g'(k)\cdot(x-k)+\int_{0}^{k} g''(y) (y-x)^{+}\,dy +\int_{k}^{\infty} g''(y) (x-y)^{+}\,dy\,.
\end{equation}
Because $x$ is the total return on one unit of wealth invested in equity, ${\rm Price}(x) =1$, and it follows
\begin{equation}\label{Eq:Price(g)}
{\rm Price}(g)=g(k)\cdot DF+g'(k)\cdot(1-k\cdot DF)+\int_{0}^{\infty} g''(y)\, {\cal O}(y,k)\,dy\,,
\end{equation}
where ${\cal O}(y,k)$ is a function comprised from vanilla option prices (both puts and calls)
\begin{equation}
{\cal O}(y,k)=\left\{
\begin{array}{ll}
{\rm price\ }{\cal P}(y){\rm\ of\ a\ put\ option\ with\ strike\ } y,& y\leq k \cr
{\rm price\ }{\cal C}(y){\rm\ of\ a\ call\ option\ with\ strike\ } y, & y> k\,.
\end{array}\right.
\end{equation}
Let us set $k:=1/DF$.
Eq.~(\ref{Eq:Price(g)}) simplifies to
\begin{equation}
{\rm Price}(g)=g(1/DF)\cdot DF+\int_{0}^{\infty} g''(y)\, {\cal O}(y,1/DF)\,dy\,.
\end{equation}
Substituting $g(x)=x^R$ we compute
\begin{equation}
{\rm Price}(x^R)=DF^{1-R}+(R^2-R)\int_{0}^{\infty} y^{R-2}\, {\cal O}(y,1/DF)\,dy\,,
\end{equation}
and
\begin{align}
\frac{\partial {\rm Price}(x^R)}{\partial R}=-DF^{1-R}\ln DF&+(2R-1)\int_{0}^{\infty} y^{R-2}\, {\cal O}(y,1/DF)\,dy\cr
                                                            &+(R^2-R)\int_{0}^{\infty} y^{R-2}\, {\cal O}(y,1/DF)\ln y\,dy\,.
\end{align}
Finally, Eq.~(\ref{Eq:ER_R}) becomes
\begin{equation}
{\rm ER}_R=\frac{-DF^{1-R}\ln DF+(2R-1)I_1(R)+(R^2-R)I_2(R)}{DF^{1-R}+(R^2-R)I_1(R)}\,,
\end{equation}
where
\begin{equation}
I_1=\int_{0}^{\infty} y^{R-2}\, {\cal O}(y,1/DF)\,dy\,, {\rm\ \ and\ \ } I_2=\int_{0}^{\infty} y^{R-2}\, {\cal O}(y,1/DF)\ln y\,dy\,.
\end{equation}
This is exactly how ${\rm ER}_R$ was computed as a function of $R$ using the option prices. Annualized values of the premium $({\rm ER}_R+\ln DF)$ were then matched to the independently reported values (see Fig.~\ref{Fig:ImpliedPremium} and Ref.~\cite{Damodaran_2014}). This was achieved by solving for $R$ using the simple bisection method. The values of $R$ implied in this way are plotted as the solid line in Fig.~\ref{Fig:Rimp}.

\subsection{Derivation of $\langle R\,\rangle_b$}\label{Sec:<R>b derivations}

Substitution of Eq.~(\ref{Eq:R}) into Eq.~(\ref{Eq:ExpectedR}) gives
\begin{equation}
\langle R\, \rangle_b\ =\ \int m \Big(\frac{b}{m}\Big)'_x\, x\, dx \ =\ \int xm\, d \Big(\frac{b}{m}\Big)\,.
\end{equation}
Integrating by parts and noticing that $xb\,\big|_0^\infty=0$, we obtain
\begin{equation}
\langle R\, \rangle_b\ =\ -\int\frac{b}{m}\, d\, (xm)\ =\ -\int\frac{b}{m}\, (m\,dx +x\,dm) \ =\ -1 - \int b\, x\, \frac{dm}{m}\,.
\end{equation}
Finally, using the notation defined by Eq.\!~(\ref{Eq:ExpectedR}), we derive
\begin{equation}\label{Eq:<R>b Analytic Appendix}
\langle R\, \rangle_b =-1-\langle\, x (\ln m)'_x\,\rangle_b\,.
\end{equation}

In the main text of the paper we illustrate this expression using the special case of a log-normal market-implied distribution. In order to see what happens in this special case we use Eq.~(\ref{Eq:LN_distr}) and derive
\begin{equation}
(\ln m)'_x \ \stackrel{\rm LN}{=}\ \Big( -\ln x -\frac{(\ln x -\mu)^2}{2\sigma^2} +{\rm const}\Big)'_x \ =\ -\frac{1}{x}-\frac{\ln x -\mu}{\sigma^2 x}\,.
\end{equation}
Substitution into Eq.~(\ref{Eq:<R>b Analytic Appendix}) gives
\begin{equation}
\langle R\, \rangle_b\ \stackrel{\rm LN}{=}\ \frac{\langle\,\ln x\,\rangle_b -\mu }{\sigma^2}=\frac{1}{2}+\frac{\langle\,\ln x\,\rangle_b -r }{\sigma^2}\,.
\end{equation}
Rearranging the terms we derive
\begin{equation}\label{Eq:<R>b Analytic explained Appendix}
 \langle\,\ln x\,\rangle_b \ \stackrel{\rm LN}{=}  r+\Big(\langle R\, \rangle_b -1/2\Big)\sigma^2 \,.
\end{equation}

\subsection{Numerical computation of $\langle R\,\rangle_P$}\label{Sec:<R>p}
Looking at Eq.~(\ref{Eq:<R>p}), computation of $\langle R\,\rangle_P$ is straightforward as long as we can compute the quantity
\begin{equation}
\Big(\ln m(x_i)\Big)'_{x_i}=\frac{m'(x_i)}{m(x_i)}\,.
\end{equation}
Let ${\cal C}(K)$ be the price of a call option with strike $K$. It is related to $m$ via the pricing formula
\begin{equation}
{\cal C}(K)=\int (x-K)^+ \,m(x)\,dx\,,
\end{equation}
which implies
\begin{equation}
m(x_i)=\frac{\partial^2{\cal C}(x_i)}{\partial x_i^2}\ \ \ {\rm and}\ \ \ m'(x_i)=\frac{\partial^3{\cal C}(x_i)}{\partial x_i^3}\,.
\end{equation}
Using market data to compute option prices as a function of strike, these quantities are easy to find by numerical differentiation. In the paper we used the straightforward finite difference approximation:
\begin{equation}
\Big(\ln m(x_i)\Big)'_{x_i}=\lim_{h\to 0}\frac{-0.5\cdot {\cal C}(x_i-2h)+{\cal C}(x_i-h)-{\cal C}(x_i+h)+0.5\cdot {\cal C}(x_i+2h)}{h\cdot\big({\cal C}(x_i-h)-2{\cal C}(x_i)+{\cal C}(x_i+h)\big)}\,.
\end{equation}

\subsection{Equilibrium market}\label{Sec:Equilibrium}
Let us consider a market of financial products which derive their value from some variable $x$. Each investor $i$ invests the amount of $w_i$ dollars by buying a product with some payoff function $F_i(x)$. Summing up across all market participants (including market makers) and assuming, for simplicity, that there are no defaults we compute
\begin{equation}\label{Eq:ClosedMarket}
\sum_iw_iF_i(x)=W/{\rm DF}\,,
\end{equation}
where $W=\sum_iw_i$ is the total amount of money invested in the market and ${\rm DF}$ is the money market discount factor.

The rest of the argument is a straightforward application (or, more accurately, an illustration) of the investor equivalence principle~\cite{Soklakov_2013b}. This principle is a thinking tool which states that for every investor one can imagine a growth-optimizing investor which chooses the same product. This is useful whenever all attention is focused on the actual actions of the investors.

Let $\beta_i(x)$ be the belief that a growth-optimizing investor must hold to buy $F_i(x)$. Mathematically,
\begin{equation}
\beta_i(x)=F_i(x)\,m(x)\,,
\end{equation}
where, assuming negligible bid-offer spreads, $m(x)$ is the market that we are looking for. Substituting this into Eq.~(\ref{Eq:ClosedMarket}) we see that $m(x)$ is given by the expression
\begin{equation}
m(x)=\frac{\rm DF}{W}\sum_i w_i\beta_i(x)\,.
\end{equation}
Note that $m(x)$ is uniquely determined by the actions of the investors. In fact, we can imagine replacing every investor with the equivalent (i.e. buying the same product) growth-optimizing investor and arrive at the same market $m(x)$.

For completeness we note that this derivation says nothing about the stability of the equilibrium. In fact, we see that each market participant exerts some influence over the market. The market dynamics in such a system would depend on the size of each investor and on how the newly discovered information spreads between the market participants of different sizes. Lux and Westerhoff argued the importance of such an interplay between heterogeneous agents~\cite{LuxWesterhoff_2009}.

\subsection{Miscellaneous notes and observations}\label{Sec:Misc}

\subsubsection{Data}
In terms of data requirements our approach differs from the consumption-based theories~\cite{MehraPrescott_1985} in two important ways. On the one hand, we do not need any data on consumption. On the other hand, we need historical records of the volatility surfaces implied by the equity options markets. The option-implied volatility surfaces are ubiquitous in the financial industry and the SPX options market is particularly well developed.

The historical data used in this paper span almost 15 years starting 17 May 2000 and ending 27 April 2015. This period contains the global financial crisis of 2007-2008 which reduced the magnitude of the historical equity premiums.\footnote{This might be partially responsible for the slight dip of risk aversion below zero for some experiments on Fig.~\ref{Fig:Rhist}, although the confidently positive values for the averages (represented by the horizontal green lines) indicate that this might also be just noise.} However, the reduction was not strong or persistent enough to remove large equity premiums across the data set used in this paper. Out of the 42 investments represented by the horizontal (green) lines on Fig.~\ref{Fig:Rhist}, the worst and the best-performing ones delivered around 2\% and 2.6\% per annum in terms of the annualized equity premium. All of these values are well above 0.35\% which was reported by Mehra and Presott as an upper limit of what can be explained by the mainstream approach~\cite{MehraPrescott_1985}.

\subsubsection{Theory}
Let us invite the reader back to the discussion around Eqs.~(\ref{Eq:<R>b Analytic}-\ref{Eq:<R>b Analytic explained}) which brings together the two separate investigations of the expected and the realized premiums. Despite the fundamental differences in meaning and different technical challenges, the two types of premiums are very similar mathematically. In essence, both of them measure the difference between some distribution (either the investor-believed $b$ or the realized $P$) and the market-implied $m$. The two premiums coincide numerically when the investor's belief is correct, i.e. when $b$ is close to $P$.

This numerical agreement between the expected and the realized premiums is a very important element within our approach because it proves that equity investments form a viable strategy which can deliver on its expectations. In order to survive, an individual strategy does not have to explain the volatility of aggregate consumption of the entire economy, but it does have to deliver on its own expectations.

We subscribe to the classic view that scientific theories must be falsifiable~\cite{Popper_1934}. To falsify our approach one would need to find a pure investment strategy which systematically underachieves its own expected performance. Long-term survival of such a strategy in a competitive economy would indeed constitute an interesting puzzle.

\end{document}